\documentclass[pra,twocolumn,showpacs]{revtex4}
\usepackage{graphicx}
\usepackage{dcolumn}
\usepackage{amsmath}
\usepackage{amssymb}
\newcommand{\bfr}{\mathbf{r}}
\newcommand{\bfu}{\mathbf{u}}

\begin{document}

\title{Density functional with full exact exchange, 
balanced nonlocality of correlation, and constraint satisfaction}
\author{John P. Perdew}
\affiliation{Department of Physics, Tulane University,
New Orleans, Louisiana 70118, USA}
\author{Viktor N. Staroverov}
\affiliation{Department of Chemistry, The University of Western Ontario,
London, Ontario N6A 5B7, Canada}
\author{Jianmin Tao}
\affiliation{Theoretical Division and CNLS, Los Alamos National Laboratory,
Los Alamos, New Mexico 87545, USA}
\author{Gustavo E. Scuseria}
\affiliation{Department of Chemistry, Rice University, Houston,
Texas 77005, USA}

\date{\today}

\begin{abstract}

We construct a nonlocal density functional approximation with
full exact exchange, while preserving the constraint-satisfaction
approach and justified error cancellations of simpler semilocal
functionals. This is achieved by interpolating between different
approximations suitable for two extreme regions of the electron
density. In a ``normal" region, the exact exchange-correlation
hole density around an electron is semilocal because its spatial
range is reduced by correlation and because it integrates
over a narrow range to $-1$. These regions are well described
by popular semilocal approximations (many of which have been
constructed nonempirically), because of proper accuracy for a
slowly-varying density or because of error cancellation between
exchange and correlation. ``Abnormal" regions, where nonlocality is
unveiled, include those in which exchange can dominate correlation
(one-electron, nonuniform high-density, and rapidly-varying limits),
and those open subsystems of fluctuating electron number over
which the exact exchange-correlation hole integrates to a value
greater than $-1$. Regions between these extremes are described
by a hybrid functional mixing exact and semilocal exchange energy
densities locally, i.e., with a mixing fraction that is a function
of position $\bfr$ and a functional of the density. Because our
mixing fraction tends to 1 in the high-density limit, we employ full
exact exchange according to the rigorous definition of the exchange
component of any exchange-correlation energy functional. Use of full
exact exchange permits the satisfaction of many exact constraints,
but the nonlocality of exchange also requires balanced nonlocality
of correlation. We find that this nonlocality can demand at least
five empirical parameters, corresponding roughly to the four kinds of
abnormal regions. Our local hybrid functional is perhaps the first
accurate fourth-rung density functional or hyper-generalized gradient
approximation, with full exact exchange, that is size-consistent in
the way that simpler functionals are. It satisfies other known exact
constraints, including exactness for all one-electron densities,
and provides an excellent fit to the 223 molecular enthalpies of
formation of the G3/99 set and the 42 reaction barrier heights of
the BH42/03 set, improving both (but especially the latter) over most
semilocal functionals and global hybrids. Exact constraints, physical
insights, and paradigm examples hopefully suppress ``overfitting".

\end{abstract}

\pacs{31.15.ej, 31.10.+z, 71.15.Mb}
\maketitle

\section{Introduction}
\label{sec:intro}

Kohn--Sham density functional
theory~\cite{Kohn:1965/PR/A1133,Book/Fiolhais:2003} is a
computationally efficient and often usefully accurate approach
to electronic structure calculation in condensed matter physics
and quantum chemistry. In this theory, the ground-state electron
density and total energy $E$ are found by self-consistent solution of
a one-electron Schr\"{o}dinger equation. Only the exchange-correlation
(xc) energy as a functional of the density needs to be approximated
in practice. This energy can always be written as
\begin{equation}
 E_\mathrm{xc}[n_\uparrow,n_\downarrow]
  = \int d^3r\, n(\bfr)\epsilon_\mathrm{xc}
 ([n_\uparrow,n_\downarrow];\bfr).  \label{eq:exc}
\end{equation}
In this equation, the energy density is the product of the
electron density $n(\bfr)=n_\uparrow(\bfr)+n_\downarrow(\bfr)$
(the sum of the spin densities) and the position-dependent
exchange-correlation energy per electron
$\epsilon_\mathrm{xc}([n_\uparrow,n_\downarrow];\bfr)$.

Local or semilocal approximations for
$\epsilon_\mathrm{xc}([n_\uparrow,n_\downarrow];\bfr)$, as defined
below, are computationally fast and conceptually simple, and can
be constructed without empiricism.  They work best for $sp$-block
atoms and their molecules and solids around equilibrium. In
real systems, full nonlocality is most needed to correct the
so-called self-interaction errors~\cite{Perdew:1981/PRB/5048},
which can be most severe in certain stretched-bond situations
that arise in transition states for chemical reactions and
in the dissociation limit (and also effectively in the
$d$- and $f$-block systems)~\cite{Perdew:1982/PRL/1691,%
Perdew:1985/Dreizler/265,Ruzsinszky:2006/JCP/194112,%
Ruzsinszky:2007/JCP/104102,Mori-Sanchez:2006/JCP/201102,%
Vydrov:2007/JCP/154109,Perdew:2007/PRA/040501,%
Ruzsinszky:2008/PRA/060502}. In this paper, we follow a conservative
approach to full nonlocality: We identify those ``normal"
regions of space in which semilocality is justified, and those
``abnormal" ones in which it is not, and then introduce full
nonlocality only to the extent that a region is abnormal. The
density functional that we construct here shows promising early
results, but, whether or not that promise is ultimately fulfilled,
the underlying insights could have lasting value.

In popular approximations,
$\epsilon_\mathrm{xc}([n_\uparrow,n_\downarrow];\bfr)$ is a
function of ingredients at position $\bfr$, and different
selections of ingredients define different rungs of a
``Jacob's ladder"~\cite{Perdew:2001/VanDoren/1} of
approximations. Accuracy tends to increase up the ladder,
while simplicity increases downwards. The first three rungs
use semilocal ingredients constructed from the density or
orbitals in an infinitesimal neighborhood of $\bfr$, and
have nonempirical~\cite{Perdew:2005/JCP/062201} (as well
as empirical) constructions.  The first rung is the local
spin density approximation (LSDA)~\cite{Kohn:1965/PR/A1133,%
Vosko:1980/CJP/1200,Perdew:1992/PRB/13244}, which uses as ingredients
$n_\uparrow(\bfr)$ and $n_\downarrow(\bfr)$ and is exact only for a
uniform or very slowly-varying electron gas. The second rung is the
generalized gradient approximation (GGA), which adds the gradients
$\nabla n_\uparrow(\bfr)$ and $\nabla n_\downarrow(\bfr)$, as in
the Perdew--Burke--Ernzerhof (PBE)~\cite{Perdew:1996/PRL/3865} and
PBEsol~\cite{Perdew:2008/PRL/136406} nonempirical constructions.
The third rung is the meta-GGA, which adds the positive
orbital kinetic energy densities $\tau_\uparrow(\bfr)$ and
$\tau_\downarrow(\bfr)$, as in the Tao--Perdew--Staroverov--Scuseria
(TPSS)~\cite{Tao:2003/PRL/146401,Staroverov:2003/JCP/12129,%
Staroverov:2004/PRB/075102} nonempirical construction. Alternatively,
the Laplacians $\nabla^2 n_\uparrow(\bfr)$ and $\nabla^2
n_\downarrow(\bfr)$ of the spin densities could be
used~\cite{Perdew:2007/PRB/155109}. Self-consistent calculations
with any of the semilocal functionals are relatively easy and
efficient, especially if the demand for a single multiplicative
effective potential is dropped at the meta-GGA level (as in
Refs.~\onlinecite{Tao:2003/PRL/146401,Staroverov:2003/JCP/12129,%
Staroverov:2004/PRB/075102}).

The fourth rung or
hyper-GGA~\cite{Perdew:2001/VanDoren/1,Cruz:1998/JPCA/4911,%
Jaramillo:2003/JCP/1068,Becke:2005/JCP/064101,Becke:2007/JCP/124108,%
Mori-Sanchez:2006/JCP/091102,Cohen:2007/JCP/034101,%
Arbuznikov:2006/JCP/204102,Kaupp:2007/JCP/194102,%
Arbuznikov:2008/JCP/214107}, with which we will be concerned here,
adds a fully nonlocal (i.e., not semilocal) ingredient,
the exact-exchange energy per
electron $\epsilon_\mathrm{x}^\mathrm{ex}$.  Unlike total energies
$E$, energy densities $n\epsilon_\mathrm{xc}$ are non-unique or
gauge-dependent. In the conventional gauge,
\begin{equation}
\epsilon_\mathrm{x}^\mathrm{ex(conv)}(\bfr) 
 = \frac{1}{2}\int d^3r' \,
  \frac{n_\mathrm{x}(\bfr,\bfr')}{|\bfr' - \bfr|},
  \label{eq:ex-conv}
\end{equation}
where
\begin{equation}
 n_\mathrm{x}(\bfr,\bfr')
  = -\sum_{\sigma}\frac{|\rho_{\sigma}(\bfr,\bfr')|^2}{n(\bfr)},
  \label{eq:xhole}
\end{equation}
is the exact-exchange hole density around an electron at $\bfr$, and
\begin{equation}
 \rho_{\sigma}(\bfr,\bfr') = \sum_i f_{i\sigma}
  \psi_{i\sigma}(\bfr) \psi_{i\sigma}^{*}(\bfr')
  \label{eq:denmat}
\end{equation}
is the Kohn--Sham one-particle density matrix for spin $\sigma$.
The $f_{i\sigma}$ are occupation numbers, typically 0 or
1 except for values in between that can arise when a Kohn--Sham
orbital $\psi_{i\sigma}(\bfr)$ is shared with another
system~\cite{Perdew:1981/PRB/5048,Perdew:1982/PRL/1691,%
Perdew:1985/Dreizler/265,Ruzsinszky:2006/JCP/194112,%
Ruzsinszky:2007/JCP/104102,Mori-Sanchez:2006/JCP/201102,%
Vydrov:2007/JCP/154109,Perdew:2007/PRA/040501,%
Ruzsinszky:2008/PRA/060502}.
The exact-exchange hole density satisfies the sum rule
\begin{eqnarray}
 \int d^3r'\, n_\mathrm{x}(\bfr,\bfr')
 & = & -1 + \sum_\sigma \sum_i f_{i\sigma}
 (1-f_{i\sigma}) \frac{|\psi_{i\sigma}(\bfr)|^2}{n(\bfr)}
 \nonumber \\
 & = & -1 + \Delta_\mathrm{x}(\bfr),
  \quad\quad (0\leq\Delta_\mathrm{x}\leq 1) \label{eq:sum-rule}
\end{eqnarray}
and the right-hand side of Eq.~(\ref{eq:sum-rule}) reduces to $-1$
when all occupation numbers are 0 or 1.
A rigorous proof of the exchange-only
equations~(\ref{eq:ex-conv})--(\ref{eq:sum-rule}) for an
open system of fluctuating electron number is given in
Ref.~\onlinecite{Perdew:2007/PRA/040501}. There is also a fifth
rung (e.g., Ref.~\onlinecite{Constantin:2008/PRL/036401}) which
adds all the occupied \textit{and} unoccupied orbitals as ingredients.
While the fourth- and fifth-rung functionals can also be implemented
fully self-consistently, tests and applications using, say, meta-GGA
orbitals (as in the present work) are easier and more efficient
and often suffice.

As we ascend the ladder, the additional ingredients
can be used to satisfy additional exact constraints on
$E_\mathrm{xc}[n_\uparrow,n_\downarrow]$ (nonempirical
approach~\cite{Perdew:2001/VanDoren/1,Perdew:2005/JCP/062201}),
or to fit data better (empirical approach), or both. As we
will see here, many additional constraints can be satisfied
on the fourth rung, including exactness for all one-electron
densities~\cite{Perdew:1981/PRB/5048,Perdew:2001/VanDoren/1,%
Jaramillo:2003/JCP/1068} and full exact
exchange~\cite{Perdew:2001/VanDoren/1}, but for the first time some
empiricism becomes unavoidable (although it can be avoided again
on the fifth rung~\cite{Constantin:2008/PRL/036401}).

There are good reasons for these particular choices of ingredients.
The local density suffices for a uniform electron gas.
The density gradients  contribute to second-order gradient
expansions~\cite{Antoniewicz:1985/PRB/6779,Ma:1968/PR/18,%
Langreth:1980/PRB/5469} for densities that vary
slowly over space, and can also satisfy other
constraints~\cite{Perdew:1996/PRL/3865}. The
positive orbital kinetic energy density is relevant to
exchange~\cite{Becke:1989/PRA/3761,Tao:2003/PRL/146401}
and in particular to its fourth-order gradient
expansion~\cite{Svendsen:1996/PRB/17402,Tao:2003/PRL/146401},
and can be used to zero out the correlation energy density in
one-electron regions~\cite{Becke:1998/JCP/2092,Tao:2003/PRL/146401}.
The Kohn--Sham orbitals needed to construct the higher-rung
ingredients are themselves fully nonlocal but implicit functionals
of the density. (Strictly, the orbitals are density functionals
when the optimum effective potential is required to be a function
of position $\bfr$. In practice, ground-state densities and energies
are not much affected when this requirement is dropped, as it often
is for computational convenience in self-consistent implementations
of meta-GGA's.)

The ladder classification is not
intended to be exclusive. Range-separated
hybrids~\cite{Leininger:1997/CPL/151,Vydrov:2006/JCP/234109,%
Vydrov:2007/JCP/154109} that use the exact-exchange hole density
of Eq.~(\ref{eq:xhole}) fall slightly above the fourth rung but
well below the fifth. Fully nonlocal ``two-point'' approximations
or six-dimensional integrals that are explicit functionals of the
density are also possible, and can provide a long-range van der
Waals interaction~\cite{Dion:2004/PRL/246401} that is still missing
from our local hybrid functional, but implicit density functionals
with a nonlocal dependence upon the orbitals are probably needed to
describe the situation in which electrons are shared between open
subsystems~\cite{Perdew:1982/PRL/1691,Perdew:1985/Dreizler/265,%
Ruzsinszky:2006/JCP/194112,Ruzsinszky:2007/JCP/104102,%
Vydrov:2007/JCP/154109,Mori-Sanchez:2006/JCP/201102,%
Perdew:2007/PRA/040501}.

There are two principal reasons why density functionals should
be constructed to satisfy known exact constraints.  The first
is idealistic: Our approximations should not needlessly violate
what we know to be true.  The second is practical:  Satisfaction
of many constraints helps the approximation to work over a wider
range of densities. For example, functionals that are constructed
only by fitting to molecular data are less accurate for solids than
functionals that satisfy solid-state-like constraints such as the
uniform gas limit~\cite{Kurth:1999/IJQC/889}, which by itself fully
defines the first or LSDA rung of the ladder.

At the second or GGA rung, the constraints are already
less constraining. There are several different sets
of constraints that can be imposed on this rung,
and some are (for this limited form) incompatible with
others~\cite{Perdew:1996/PRL/3865,Perdew:2008/PRL/136406}. Even for
a chosen set of constraints, there may be quite different ways in
which they can be satisfied~\cite{Csonka:2007/JCP/244107}. Thus,
the nonempirical GGAs are in practice guided by some additional
physical postulate. For the PBE GGA, this guiding postulate
is the sharp (atomic-like) cutoff of the spurious long-range
part of the gradient expansion for the exchange-correlation
hole~\cite{Perdew:1996/PRB/16533}. For the PBEsol GGA for solids,
it is restoration of the gradient expansion for the exchange
energy over a wide range of slowly- or moderately-varying
densities~\cite{Perdew:2008/PRL/136406}.

We close this introduction by summarizing some of the successes
and failures of the TPSS meta-GGA, the nonempirical functional
on the highest semilocal rung of the ladder. For the standard
enthalpies of formation of the 223 molecules of the G3/99 test
set~\cite{Curtiss:2000/JCP/7374} (including COF$_2$), the mean
absolute error (MAE) computed self-consistently using the fully
uncontracted 6-311++G(3df,3pd) basis set is only 6.5 kcal/mol
when TPSS exchange is combined with TPSS correlation. The MAE
increases to 30.3 kcal/mol when full exact exchange is combined
with TPSS correlation, and to 211.0 kcal/mol for exact exchange
and no correlation (i.e., Hartree--Fock).
Thus, atoms and molecules at equilibrium, with
integer electron numbers, are predominantly normal systems in which
the errors of semilocal exchange and semilocal correlation tend
to cancel. This does \textit{not} mean that semilocal functionals
always predict realistic binding energy curves. In fact, they do so
only when the fragments have integer electron numbers. Asymmetric
molecules, when described by semilocal functionals, often dissociate
to fragments of spurious fractional charge, for which the energy
is seriously too low (Ref.~\onlinecite{Ruzsinszky:2006/JCP/194112}
and references therein). Even when the fractional charge is real,
as in the dissociation of X$_2^{+}$, the self-interaction error
of semilocal functionals makes the energy seriously too low
(Ref.~\onlinecite{Ruzsinszky:2007/JCP/104102}, and references
therein) at the dissociation limit X$^{+1/2}\cdots$X$^{+1/2}$,
where it ought to be equal to that of X$^{0}\cdots$X$^{+1}$. The
combination of stretched bonds and noninteger average (hence
fluctuating) electron numbers on the fragments makes for a highly
abnormal valence region. A milder version of the same abnormality
is found for the moderately stretched bonds in the transition state
of a chemical reaction, where semilocal functionals predict reaction
barriers that are too low or even negative.

\section{Local hybrid form of the hyper-GGA}
\label{sec:lh}

The earliest hyper-GGAs were the global hybrid
functionals~\cite{Becke:1993/JCP/5648,Perdew:1996/JCP/9982,%
Zhao:2006/JCTC/364,Staroverov:2003/JCP/12129},
for which the simplest form is
\begin{equation}
 E_\mathrm{xc}^\mathrm{gh} = aE_\mathrm{x}^\mathrm{ex}
 + (1-a)E_\mathrm{x}^\mathrm{sl} + E_\mathrm{c}^\mathrm{sl}.
  \label{eq:gh}
\end{equation}
Here $E_\mathrm{x}^\mathrm{sl}$ and $E_\mathrm{c}^\mathrm{sl}$
are semilocal exchange and correlation functionals,
$E_\mathrm{x}^\mathrm{ex}$ is the exact-exchange energy, and
$a$ is a universal, position-independent empirical mixing
fraction between 0 and 1. While semilocal approximations
($a=0$) typically overestimate the atomization energies of
molecules~\cite{Curtiss:1997/JCP/1063,Staroverov:2003/JCP/12129}
and underestimate the energy barriers to chemical
reactions~\cite{Zhao:2004/JPCA/2715}, the use of exact
or Hartree--Fock exchange ($a=1$) makes opposite errors.
Fitting Eq.~(\ref{eq:gh}) to atomization energies often
gives $a\approx 0.2$, while the reaction barrier heights typically
require $a\approx 0.5$. For some systems and properties, even
$a\approx 0.2$ is too large~\cite{Furche:2006/JCP/044103}.
The smallness of $a$ indicates that semilocal exchange is
more compatible with semilocal correlation than is the exact,
fully nonlocal exchange. The global hybrid functionals
are perhaps the most popular functionals in modern quantum
chemistry. (Screened hybrid functionals~\cite{Heyd:2003/JCP/8207}
are also gaining popularity for solid-state
calculations~\cite{Heyd:2004/JCP/1187,Paier:2006/JCP/154709}).
They often improve upon the semilocal functionals, at the
sometimes-acceptable additional cost of a Hartree--Fock calculation,
and the improvement can be pushed further by the addition of further
empirical parameters~\cite{Zhao:2006/JCTC/364}.
However, they satisfy no exact constraint beyond those satisfied
by the underlying semilocal $E_\mathrm{x}^\mathrm{sl}$.

A natural generalization to Eq.~(\ref{eq:gh}) is the local
hybrid functional
\begin{eqnarray}
\epsilon_\mathrm{xc}^\mathrm{lh}(\bfr)
 & = & a(\bfr)\epsilon_\mathrm{x}^\mathrm{ex}(\bfr) +
 [1 - a(\bfr)] \epsilon_\mathrm{x}^\mathrm{sl}(\bfr)
 + \epsilon_\mathrm{c}^\mathrm{sl}(\bfr) \nonumber \\
 & = & \epsilon_\mathrm{x}^\mathrm{ex}(\bfr) +
 [1 - a(\bfr)][\epsilon_\mathrm{x}^\mathrm{sl}(\bfr)
 - \epsilon_\mathrm{x}^\mathrm{ex}(\bfr)]
 + \epsilon_\mathrm{c}^\mathrm{sl}(\bfr),   \label{eq:lh}
\end{eqnarray}
where $0\leq a(\bfr)\leq 1$. Eq.~(\ref{eq:lh}) was first
proposed in Ref.~\onlinecite{Cruz:1998/JPCA/4911},
without a form for $a(\bfr)$. Forms were proposed
in Ref.~\onlinecite{Perdew:2001/VanDoren/1} and
Ref.~\onlinecite{Jaramillo:2003/JCP/1068} (where the term
``local hybrid" was coined), but those forms did not achieve
high accuracy for equilibrium properties of molecules.

The choice
$\epsilon_\mathrm{xc}^\mathrm{sl}=\epsilon_\mathrm{xc}^\mathrm{TPSS}$
and $a(\bfr)=1$ in Eq.~(\ref{eq:lh}) satisfies nearly all
exact constraints that a hyper-GGA can satisfy, but yields
very poor atomization energies of molecules (as shown in
Ref.~\onlinecite{Perdew:2001/VanDoren/1}, for example) because
it misses the delicate and helpful error cancellation between
semilocal exchange and semilocal correlation that typically occurs
(because the exact xc-hole is deeper and more short-ranged than
the exact x-hole) in ``normal" regions of space (as defined in
section~\ref{sec:normal}).  From this fact, it is already clear
that the known exact constraints do not tell us how much full
nonlocality is needed in correlation for compatibility with exact
exchange.  The imbalance between nonlocal approximate exchange
and semilocal approximate correlation may also be responsible
for the inaccuracies of the Perdew--Zunger self-interaction
correction~\cite{Perdew:1981/PRB/5048} and of the local hybrid
functional of Ref.~\onlinecite{Jaramillo:2003/JCP/1068}, when
applied to atomization energies.

As envisioned in Ref.~\onlinecite{Perdew:2001/VanDoren/1}
and discussed below in section~\ref{subsec:4.1}, a hyper-GGA
has full exact exchange if it satisfies the exact
constraint~\cite{Levy:1991/PRA/4637} (for any system with a
nondegenerate Kohn--Sham noninteracting ground state)
\begin{equation}
 \lim_{\lambda\to\infty} \frac{E_\mathrm{xc}[n_\lambda]}
 {E_\mathrm{x}^\mathrm{ex}[n_\lambda]} = 1   \label{eq:E-rat-lim}
\end{equation}
under uniform density scaling
\begin{equation}
 n(\bfr)\to n_\lambda(\bfr)=\lambda^3 n(\lambda\bfr)
 \label{eq:uds}
\end{equation}
to the high-density limit $\lambda\to\infty$. (Note that
$n(\bfr)$ and $n_\lambda(\bfr)$ have the same electron
number). Such a functional will automatically satisfy all
exact constraints on exchange alone. Of several developed
hyper-GGAs~\cite{Becke:2005/JCP/064101,Becke:2007/JCP/124108,%
Mori-Sanchez:2006/JCP/091102,Cohen:2007/JCP/034101,%
Cohen:2007/JCP/191109,Arbuznikov:2006/JCP/204102,Kaupp:2007/JCP/194102},
the only one that seems to have this property is that of
Refs.~\onlinecite{Mori-Sanchez:2006/JCP/091102,Cohen:2007/JCP/034101},
and~\onlinecite{Cohen:2007/JCP/191109}, for which the correlation
energy is not exactly size-consistent. Our size-consistent local
hybrid has full exact exchange because its mixing fraction
$a(\bfr)$ tends rapidly enough to 1 in the high-density
limit, while $\epsilon_\mathrm{x}^\mathrm{ex}\sim\lambda$,
$\epsilon_\mathrm{x}^\mathrm{sl}\sim\lambda$, and
$\epsilon_\mathrm{c}^\mathrm{sl}\sim\lambda^0$ in this limit.
Formally, the exchange part of any $E_\mathrm{xc}[n]$
is~\cite{Levy:1991/PRA/4637}
\begin{equation}
 E_\mathrm{x}[n] = \lim_{\lambda\to\infty}
 \frac{E_\mathrm{xc}[n_\lambda]}{\lambda}.  \label{eq:Ex-lambda}
\end{equation}
Equation (10) is the only non-arbitrary way to define the exchange
component of a density functional for the exchange-correlation energy.

Thus, the three terms in the last line of Eq.~(\ref{eq:lh}) represent
respectively exact exchange, the fully nonlocal part of correlation
including ``left-right"~\cite{Handy:2001/MP/403} static or
near-degeneracy correlation important in molecules, and the semilocal
dynamic correlation. Instead of thinking of $a(\bfr)$ as the fraction
of exact exchange mixed locally with semilocal exchange, it is more
correct to think of $[1-a(\bfr)]$ as the fraction of the nonlocal
energy difference $[\epsilon_\mathrm{x}^\mathrm{sl}(\bfr)
-\epsilon_\mathrm{x}^\mathrm{ex}(\bfr)]$
added to exact exchange plus semilocal correlation. The quantity
$[\epsilon_\mathrm{x}^\mathrm{sl}(\bfr)-\epsilon_\mathrm{x}^\mathrm{ex}(\bfr)]$
is the nonlocality error (and arguably often the self-interaction
error) of the semilocal $\epsilon_\mathrm{x}^\mathrm{sl}$.
In a normal region $(a\approx 0)$, it provides an estimate of
the nonlocal part of the correlation energy (such as the static
correlation).  In an abnormal region, the semilocal functionals often
overestimate~\cite{Perdew:2007/PRA/040501} the nonlocal correction
to correlation, and their estimate must be scaled back. Our $a(\bfr)$
is mnemonic for and measures the ``abnormality" of a region of space.

We have recently argued~\cite{Perdew:2007/PRA/040501}
that Eq.~(\ref{eq:lh}) could be useful not only for molecules
and solids near equilibrium, but even for the more challenging
problem of stretched bonds between open subsystems with noninteger
average electron number.  In such subsystems, in the limit of
infinite bond stretch, the  total energy variation with electron
number between adjacent integers is concave downward for exact
exchange~\cite{Perdew:2007/PRA/040501}, concave upward for semilocal
exchange or exchange-correlation~\cite{Perdew:1985/Dreizler/265},
and linear in an exact description~\cite{Perdew:1982/PRL/1691,%
Perdew:1985/Dreizler/265}, justifying the local-hybrid
mixing of Eq.~(\ref{eq:lh}). To the extent that a functional
mimics the exact linearity, it is said to be ``many-electron
self-interaction-free''~\cite{Ruzsinszky:2006/JCP/194112,%
Ruzsinszky:2007/JCP/104102,Mori-Sanchez:2006/JCP/201102}.
We have also discussed~\cite{Perdew:2007/PRA/040501} in a
general way how $a(\bfr)$ should be much less than 1 in
normal regions where $\epsilon_\mathrm{xc}^\mathrm{sl}$ can
be accurate, but should approach 1 in many abnormal regions
where $\epsilon_\mathrm{xc}^\mathrm{sl}$ cannot be accurate.
We shall construct $a(\bfr)$ in sections~\ref{sec:normal}
and~\ref{sec:mixing}. This construction will necessarily be
complicated, because we aim to satisfy as many exact constraints as
possible. Our construction will explain why the more complicated
hyper-GGAs~\cite{Becke:2005/JCP/064101,Becke:2007/JCP/124108,%
Mori-Sanchez:2006/JCP/091102,Cohen:2007/JCP/034101}
(including ours) require typically four or five
empirical parameters to balance the full nonlocality of
correlation with that of exchange.  (The hyper-GGAs of
Refs.~\onlinecite{Becke:2005/JCP/064101,Becke:2007/JCP/124108,%
Mori-Sanchez:2006/JCP/091102,Cohen:2007/JCP/034101} have additional
empirical parameters in their semilocal parts, while our hyper-GGA
does not.)

To satisfy as many constraints as possible, we will take
\begin{equation}
 \epsilon_\mathrm{x}^\mathrm{sl}(\bfr)
 = \epsilon_\mathrm{x}^\mathrm{TPSS}(\bfr), \quad\quad
 \epsilon_\mathrm{c}^\mathrm{sl}(\bfr)
 = \epsilon_\mathrm{c}^\mathrm{TPSS}(\bfr).  \label{eq:epsilons}
\end{equation}
The TPSS meta-GGA~\cite{Tao:2003/PRL/146401,%
Staroverov:2003/JCP/12129,Staroverov:2004/PRB/075102}
is constructed nonempirically to satisfy those exact
constraints that a meta-GGA can, including the constraint
$\epsilon_\mathrm{c}(\bfr)=0$ in one-electron regions.

A complication arises because of the non-uniqueness of the
exchange energy density: many different energy densities can
integrate to the same energy. In Eq.~(\ref{eq:lh}), we want
$\epsilon_\mathrm{x}^\mathrm{ex}(\bfr)$ to be as similar to
$\epsilon_\mathrm{x}^\mathrm{sl}(\bfr)$ as it can be, so we
write~\cite{Tao:2008/PRA/012509}
\begin{equation}
 n(\bfr)\epsilon_\mathrm{x}^\mathrm{ex}(\bfr)
 = n(\bfr)\epsilon_\mathrm{x}^\mathrm{ex(conv)}(\bfr)
 + \nabla\cdot[U(\bfr)\nabla V(\bfr)],  \label{eq:exg}
\end{equation}
where $\epsilon_\mathrm{x}^\mathrm{ex(conv)}$ is the conventional
exact-exchange energy per electron from Eq.~(\ref{eq:ex-conv}).
The functions $U(\bfr)$ and $V(\bfr)$ are constructed from hyper-GGA
ingredients in Ref.~\onlinecite{Tao:2008/PRA/012509}. The divergence
of the vector field $U\nabla V$ in Eq.~(\ref{eq:exg})
integrates over $\bfr$ to zero. The divergence term is largest in
regions where the density is dominated by a single orbital shape,
such as one-electron and tail regions; omitting it increases the
errors of our local hybrid functional quantitatively but not
qualitatively. The global hybrids of Eq.~(\ref{eq:gh}) are of
course invariant under any choice of gauge. 

We have argued that the empirical and nonempirical approaches to
density functional approximation must flow together at the hyper-GGA
level, as they need not on the three lower rungs of the ladder.
On this fourth rung, the world is almost turned upside down: Up
to a point, building more physics into a hyper-GGA demands more
and not fewer empirical parameters.  (A similar situation arises
in the construction of explicit density functionals for the
orbital kinetic energy~\cite{Perdew:2007/PRB/155109}.) 

Simple one-parameter local hybrid functionals,
satisfying few or no exact constraints beyond those of the
lower-rung functionals, have been constructed by Kaupp and
collaborators~\cite{Arbuznikov:2006/JCP/204102,Kaupp:2007/JCP/194102}.
More complicated functionals with four or five empirical parameters
have been developed using all three standard physical
approaches to density functional approximation:

(a) Modeling~\cite{Becke:1989/PRA/3761,Perdew:1996/PRB/16533}
the exchange-correlation hole around an electron, including the
static-correlation part that arises in stretched-bond H$_2$, leads to
Becke's hyper-GGA~\cite{Becke:2005/JCP/064101,Becke:2007/JCP/124108}.
Because Becke starts from a system that has an \textit{exact}
degeneracy of its Kohn--Sham noninteracting ground state, he is able
to avoid some of the symmetry breaking that persists in our and other
hyper-GGAs, but at the cost of violating the exact constraint
of Eq.~(\ref{eq:E-rat-lim}) for all ground states without such an
unusual degeneracy.  For any system in which Becke's hyper-GGA
and ours predict a non-zero static correlation energy, the ratio
of static correlation to exchange energies under uniform density
scaling to the high-density limit will remain unchanged for Becke's
hyper-GGA while it will tend to zero for ours; the latter behavior
is the correct one in the normal case where the static correlation
in the unscaled system arises from a near- but not exact degeneracy.

(b) Modeling~\cite{Becke:1993/JCP/5648,Perdew:1996/JCP/9982,%
Seidl:2000/PRL/5070} the adiabatic
connection~\cite{Langreth:1975/SSC/1425} between
the Kohn--Sham Slater determinant and the true
interacting wave function leads to the hyper-GGA of Yang and
collaborators~\cite{Mori-Sanchez:2006/JCP/091102,Cohen:2007/JCP/034101},
but with a loss of size consistency (in the weak sense in which simpler
functionals are size-consistent). Without size consistency,
the energy of a subsystem (e.g., an atom or molecule) can
depend upon the presence of another (e.g., a metal surface)
even in the limit of infinite separation and even without
charge transfer between the subsystems.  Size consistency
can be preserved by using energy densities, as here and in
Refs.~\onlinecite{Becke:2005/JCP/064101,Becke:2007/JCP/124108}.
It is lost when total energies are used nonlinearly.

(c) Satisfaction~\cite{Wilson:1990/PRB/12930,Perdew:1996/PRL/3865,%
Tao:2003/PRL/146401} of exact constraints on
$E_\mathrm{xc}[n_\uparrow,n_\downarrow]$ leads to our local hybrid.

Hyper-GGAs have also been constructed by fitting
a large number of empirical parameters to chemical
data sets (e.g., Refs.~\onlinecite{Zhao:2006/JCTC/364}
and~\onlinecite{Zhao:2006/JPCA/13126}), while satisfying
relatively few exact constraints. In particular, the M06-HF
hyper-GGA of Zhao and Truhlar~\cite{Zhao:2006/JPCA/13126} is
claimed to have ``full Hartree-Fock exchange", but like Becke's
hyper-GGA~\cite{Becke:2005/JCP/064101,Becke:2007/JCP/124108}
it contains in addition other terms (semi-local ones in this
case) which improperly scale like exchange in the high-density
limit. Plots~\cite{Zhao:2006/JCTC/364} of the enhancement factors
of the semi-local parts of the parent functional M05-2X display
wiggles and other anomalous behaviors suggesting more parameters
and fewer constraints than are needed.

Our hyper-GGA work here is distinguished from  that of others in
two principal ways: (i) We aim to understand fully and build in the
error cancellation~\cite{Perdew:2007/PRA/040501} between semilocal
exchange and semilocal correlation that is responsible for much of
the success of simpler functionals. (ii) We aim to satisfy nearly
all the known exact constraints that a hyper-GGA can, in order to
test and apply this exact knowledge and to make our functional
useful even in systems and situations very different from those
where we have fitted or tested it. One example of such a system
is jellium uniaxially compressed toward the true two-dimensional
limit~\cite{Constantin:2008/PRL/016406}. For this challenging
problem of dimensional cross-over, our local hybrid functional
is significantly better than LSDA, PBE GGA, or TPSS meta-GGA,
although not as good as the fifth-rung functional described in
Ref.~\onlinecite{Constantin:2008/PRL/036401}.

\section{Why the exact exchange-correlation hole is semilocal
in normal regions}
\label{sec:normal}

The exact exchange-correlation energy can be
written~\cite{Langreth:1975/SSC/1425,Gunnarsson:1976/PRB/4274} in
the form of Eq.~(\ref{eq:exc}), where in the conventional gauge
\begin{eqnarray}
 \epsilon_\mathrm{xc}^\mathrm{ex(conv)}(\bfr)
 & = & \frac{1}{2} \int d^3r' \int_0^1 d\alpha\,
 \frac{n_\mathrm{xc}^\mathrm{ex,\alpha}(\bfr,\bfr')}{|\bfr'-\bfr|} \nonumber \\
 & & = \frac{1}{2} \int_0^\infty du\,
  4\pi u \bar{n}_\mathrm{xc}^\mathrm{ex}(\bfr,u).  \label{eq:nex}
\end{eqnarray}
Here $\bar{n}_\mathrm{xc}^\mathrm{ex}(\bfr,u)$ is
the spherical average over the direction of $\bfu=\bfr'-\bfr$
of the average over coupling constant $\alpha$ of
$n_\mathrm{xc}^\mathrm{ex,\alpha}(\bfr,\bfr')$, where
the density is held fixed as $\alpha$ varies from 0 to 1
in the Coulomb interaction $\alpha/|\bfr'-\bfr|$, and
$n_\mathrm{xc}^\mathrm{ex,\alpha=0}(\bfr,\bfr')
=n_\mathrm{x}^\mathrm{ex}(\bfr,\bfr')$. The exact xc-hole obeys
the sum rule
\begin{equation}
 \int_0^\infty du\, 4\pi u^2
  \bar{n}_\mathrm{xc}^\mathrm{ex}(\bfr,u) 
 = -1 + \Delta_\mathrm{xc}(\bfr),
 \quad\quad (0\leq\Delta_\mathrm{xc}\leq 1),
  \label{eq:xc-hole-sum}
\end{equation}
where $\Delta_\mathrm{xc}$ vanishes if the
integrated electron number in the system does not
fluctuate~\cite{Perdew:1985/Dreizler/265,Perdew:2007/PRA/040501}.
In the absence of exact degeneracy of the Kohn--Sham noninteracting
system, the $\alpha=0$ limit of $-1+\Delta_\mathrm{xc}$ is shown
on the extreme right-hand side of Eq.~(\ref{eq:sum-rule}). For
the slowly-varying densities on which the semilocal
approximations are based, $\Delta_\mathrm{xc}=\Delta_\mathrm{x}=0$.
For the LSDA, $\bar{n}_\mathrm{xc}^\mathrm{LSDA}(\bfr,u)
=\bar{n}_\mathrm{xc}^\mathrm{unif}
(n_\uparrow(\bfr),n_\downarrow(\bfr);u)$; for the other
semilocal approximations, the hole can be either modeled from
the start~\cite{Perdew:1996/PRB/16533,Becke:1989/PRA/3761}
or reverse-engineered~\cite{Ernzerhof:1998/JCP/3313,%
Constantin:2006/PRB/205104}.

Gunnarsson and Lundqvist~\cite{Gunnarsson:1976/PRB/4274} gave
an argument for the success of LSDA and hence of other semilocal
approximations: These approximations satisfy the correct sum rule
(when $\Delta_\mathrm{xc}=0$), which is by Eq.~(\ref{eq:xc-hole-sum})
the second moment of the hole density.  Then the energy per
electron, which is by Eq.~(\ref{eq:nex}) the first moment of
the hole density, should not be too wrong.  This argument was
expanded by Burke, Perdew and Ernzerhof~\cite{Burke:1998/JCP/3760}
who argued that the small-$u$ behavior (through order $u^2$)
of $\bar{n}_\mathrm{x}^\mathrm{ex}(\bfr,u)$ is determined exactly
by semilocal information, and the small-$u$ behavior (through
order $|u|$) of $\bar{n}_\mathrm{c}^\mathrm{ex}(\bfr,u)$ is
often determined approximately by local information.  They also
argued that the system average arising in Eq.~(\ref{eq:exc})
unweights regions of space in which the small-$u$ behavior of
$\bar{n}_\mathrm{xc}^\mathrm{ex}(\bfr,u)$ is not so well described
by semilocal information. This system average also eliminates the
differences between different energy-density gauges.

This argument is easily extended to explain why the semilocal
approximations usually work better for exchange and correlation
together than for either separately: Correlation makes
$\bar{n}_\mathrm{xc}^\mathrm{ex}(\bfr,u)$ deeper at small $u$,
and more short-ranged, than $\bar{n}_\mathrm{x}^\mathrm{ex}(\bfr,u)$.
A deep, narrow xc-hole density, integrating to $-1$ and
having a semilocal small-$u$ behavior, is semilocal.

We define a normal region of space~\cite{Perdew:2007/PRA/040501}
as one in which $\bar{n}_\mathrm{xc}^\mathrm{ex}(\bfr,u)$ is
modeled reasonably well by a semilocal approximation, at least
after the system average over that region.  The electron density in
a normal region is either reasonably slowly-varying, or reasonably
low and many-electron-like, or both. Only at sufficiently low and
many-electron-like density is correlation comparable in strength to
exchange, making possible a useful error cancellation between them.
In a normal region, we want $a(\bfr)\approx 0$ in Eq.~(\ref{eq:lh})
to take advantage of the proper accuracy of semilocal approximations
(for a slowly-varying density) or of the error cancellation between
semilocal exchange and semilocal correlation.

Semilocal information $(n,\nabla n,\tau)$ is enough to tell us
whether a density is slowly-varying, low, or many-electron-like.
We shall take advantage of this in section~\ref{subsec:4.1}. The
fully nonlocal information in $\epsilon_\mathrm{x}^\mathrm{ex}$
of Eqs.~(\ref{eq:ex-conv}) and~(\ref{eq:exg}) can tell us how
big $\Delta_\mathrm{x}$ is. Exact or near degeneracies can make
$\Delta_\mathrm{xc}$ small or zero even when $\Delta_\mathrm{x}$
is not small or zero (as in infinitely-stretched spin-unpolarized
H$_2$), but in those cases spin-symmetry breaking can still rescue
the semilocal approximations by restoring the correct small-$u$
behavior~\cite{Perdew:1995/PRA/4531}.  We shall make use of this
in section~\ref{subsec:4.2}.

In a normal region, we will take $a(\bfr)$ in Eq.~(\ref{eq:lh}) to
be small or zero. Conversely, an abnormal region is one in which we
have no clear reason to keep $a(\bfr)$ small, and can let $a(\bfr)$
approach 1.

By the way, we can write the spherically-averaged
exchange-correlation hole that yields our local hybrid functional
of Eq.~(\ref{eq:lh}) as
\begin{eqnarray}
 \bar{n}_\mathrm{xc}^\mathrm{lh}(\bfr,u)
 & = & \bar{n}_\mathrm{x}^\mathrm{ex}(\bfr,u)
 + [1-a(\bfr)][\bar{n}_\mathrm{x}^\mathrm{TPSS}(\bfr,u)
  - \bar{n}_\mathrm{x}^\mathrm{ex}(\bfr,u)]   \nonumber \\
 & & + \bar{n}_\mathrm{c}^\mathrm{TPSS}(\bfr,u).  \label{eq:nxc}
\end{eqnarray}
The TPSS hole is known~\cite{Constantin:2006/PRB/205104}.
Note that the middle term on the right of Eq.~(\ref{eq:nxc})
is purely long range in $u$ [and of infinite range when
$\bar{n}_\mathrm{x}^\mathrm{ex}(\bfr,u)$ is], confirming our
interpretation of this term as the nonlocal or static correlation.

\section{Mixing exact exchange in abnormal regions where the hole
is fully nonlocal}
\label{sec:mixing}

Our hyper-GGA is the local hybrid functional of Eqs.~(\ref{eq:lh}),
(\ref{eq:E-rat-lim}), and (\ref{eq:epsilons}). The motivation
for this form was presented in section~\ref{sec:lh}, and
the idea behind the local exact-exchange mixing fraction
$a(\bfr)$ was explained in section~\ref{sec:normal}
and Ref.~\onlinecite{Perdew:2007/PRA/040501}: We want
to satisfy essentially all possible exact constraints
on $E_\mathrm{xc}[n_\uparrow,n_\downarrow]$, including
Eq.~(\ref{eq:E-rat-lim}) which guarantees that our hyper-GGA has full
exact exchange, while preserving the proper accuracy or the error
cancellation between semilocal exchange and semilocal correlation
that occurs in normal regions

Thus we want $a(\bfr)\approx 0$ in a normal region, and
$a(\bfr)\approx 1$ in strongly abnormal regions. An abnormal region
is one in which the density is (i) one-electron-like, rapidly-varying
over space, or nonuniform and high, or (ii) strongly fluctuating in
electron number.  We will define an $a_1(\bfr)$ and an $a_2(\bfr)$
in subsections~\ref{subsec:4.1} and \ref{subsec:4.2}, to
identify an abnormal region according to conditions (i) or (ii),
respectively.  We will then combine these into a single $a(\bfr)$
in subsection~\ref{subsec:4.3}.

\subsection{Abnormal regions where exchange can dominate:
one-electron, rapidly-varying, and nonuniform high-density regions}
\label{subsec:4.1}

As discussed earlier, no error cancellation between exchange and
correlation can be expected in abnormal regions where the density
is too one-electron-like, too rapidly- varying  over space, or too
high, so there is no reason not to use full exact exchange there.
In regions where the density is slowly-varying over space, we might
accurately use either full exact exchange $(a=1)$ or semilocal exchange
$(a=0)$, but semilocal exchange is computationally preferable because
it avoids the need to integrate over the long tail of the exact
exchange hole.

Consider the interesting density- and position-dependent variable
\begin{equation}
 u = \frac{\epsilon_\mathrm{c}^\mathrm{GL2TPSS}}
 {\epsilon_\mathrm{x}^\mathrm{LSD}},   \label{eq:u}
\end{equation}
where
\begin{equation}
 \epsilon_\mathrm{c}^\mathrm{GL2TPSS}([n_\uparrow,n_\downarrow]; \bfr)
 = \lim_{\lambda\to\infty} \epsilon_\mathrm{c}^\mathrm{TPSS}
 ([n_{\uparrow\lambda},n_{\downarrow\lambda}];\lambda^{-1}\bfr)
  \label{eq:GL2TPSS-def}
\end{equation}
is the G\"{o}rling--Levy~\cite{Gorling:1993/PRB/13105}
second-order or high-density limit of the TPSS correlation
energy per electron at position $\bfr$. The latter
tends~\cite{Perdew:1996/PRL/3865,Tao:2003/PRL/146401,%
Perdew:1999/PRL/2544} to a negative finite limiting function of
$\lambda\bfr$ as $\lambda\to\infty$ under the uniform density
scaling of Eq.~(\ref{eq:uds}). The $\lambda^{-1}$ factor in
Eq.~(\ref{eq:GL2TPSS-def}) then restores the length scale of
the original density $n(\bfr)$. An explicit formula for
$\epsilon_\mathrm{c}^\mathrm{GL2TPSS}$ is given
in Appendix~\ref{app:A}. Moreover,
\begin{equation}
 \epsilon_\mathrm{x}^\mathrm{LSD}
 ([n_{\uparrow},n_{\downarrow}];\bfr)
 = -\frac{3}{4}\left(\frac{3}{\pi}\right)^{1/3} \frac{(2n_\uparrow)^{4/3}
  + (2n_\downarrow)^{4/3}}{2(n_\uparrow+n_\downarrow)}
\end{equation}
is the LSD exchange energy per electron at $\bfr$.  Note that
$u$ vanishes in the one-electron and rapidly-varying limits
(because $\epsilon_\mathrm{c}^\mathrm{GL2TPSS}$ does),
and $u$ also vanishes in the high-density limit (because
$1/\epsilon_\mathrm{x}^\mathrm{LSD}$ does).

\begin{figure}
\includegraphics[width=\columnwidth]{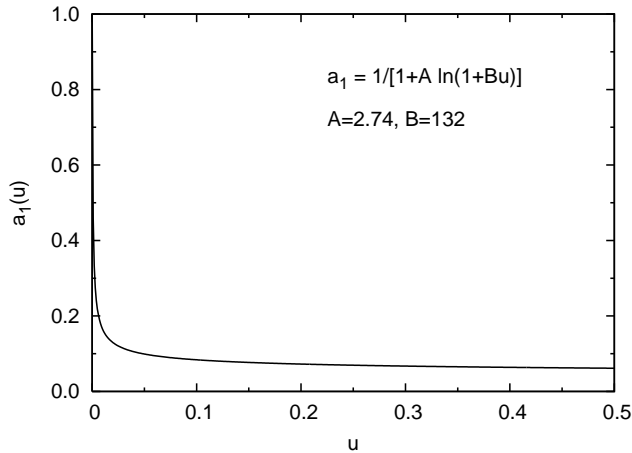}
\caption{\label{fig:a1}
Mixing fraction $a_1(u)$ of Eq.~(\ref{eq:a1}).
The values of parameters $A$ and $B$ were fitted as explained
in Section~\ref{sec:fitting}.
}
\end{figure}

Now we choose a mixing fraction $a_1(\bfr)$ as a function of $u$
which falls monotonically from 1 at $u=0$ to 0 as $u\to\infty$.
Experience with global hybrid functionals (in which $a$ is
independent of $\bfr$) suggests that $a_1$ might be a weak function
of $u$ over a wide range of $u$. Thus we try
\begin{equation}
 a_1 = \frac{1}{1+A\ln(1+Bu)}, \label{eq:a1}
\end{equation}
where $A$ and $B$ are positive empirical parameters.
This function is plotted in Fig.~\ref{fig:a1}.

In Eq.~(\ref{eq:lh}), the first term on the extreme right is the
exact-exchange energy per electron, while the sum of the middle
and last terms is the local hybrid correlation energy per electron
$\epsilon_\mathrm{c}^\mathrm{lh}$.  We will now verify that many
of the exact constraints satisfied by TPSS correlation are also
satisfied by our local hybrid correlation.

For a one-electron density, properly
$\epsilon_\mathrm{c}^\mathrm{TPSS}=0$~\cite{Tao:2003/PRL/146401,%
Perdew:1999/PRL/2544}, so $\epsilon_\mathrm{c}^\mathrm{GL2TPSS}=0$,
$u=0$, and $a_1=1$. Thus properly $\epsilon_\mathrm{c}^\mathrm{lh}=0$.
Unlike the semilocal functionals, our local hybrid is exact for all
one-electron densities (H, H$_2^{+}$, etc.)

In the rapidly-varying limit, in which
\begin{equation}
 s = \frac{|\nabla n|}{2(3\pi^2)^{1/3}n^{4/3}} \to \infty,
  \label{eq:s}
\end{equation}
$\epsilon_\mathrm{c}^\mathrm{TPSS}\to 0$~\cite{Perdew:1996/PRL/3865,%
Tao:2003/PRL/146401,Perdew:1999/PRL/2544} so
$\epsilon_\mathrm{c}^\mathrm{GL2TPSS}\to 0$, $u\to 0$, and $a_1\to 1$.
Thus, $\epsilon_\mathrm{c}^\mathrm{lh}\to 0$. In the exponential
tail of an electron density, this is the correct behavior.
Under one-dimensional density scaling to the true two-dimensional
limit~\cite{Constantin:2008/PRL/016406,Pollack:2000/JPCM/1239,%
Gorling:1992/PRA/1509}, $\epsilon_\mathrm{c}^\mathrm{lh}$ is correctly
finite, although in this case the limiting function should properly
be negative and not zero.

In the nonuniform high-density limit [$\lambda\to\infty$
under the uniform scaling of Eq.~(\ref{eq:uds})], in which
$1/\epsilon_\mathrm{x}^\mathrm{LSD}\to 0$, we find $u\to 0$ and
$a_1\to 1-ABu$. Then
\begin{equation}
 \epsilon_\mathrm{c}^\mathrm{lh} \to
 AB \epsilon_\mathrm{c}^\mathrm{GL2TPSS} \left(
 \frac{\epsilon_\mathrm{x}^\mathrm{TPSS}-\epsilon_\mathrm{x}^\mathrm{ex}}
 {\epsilon_\mathrm{x}^\mathrm{LSDA}} \right)
 + \epsilon_\mathrm{c}^\mathrm{GL2TPSS}.  \label{eq:eclh}
\end{equation}
The limit on the right of Eq.~(\ref{eq:eclh}) is
a function of $\lambda\bfr$, but one that does not
otherwise scale with $\lambda$. This is properly a finite
limit~\cite{Gorling:1993/PRB/13105} (if not always a properly
negative one), implying Eq.~(\ref{eq:E-rat-lim}) and demonstrating
that our local hybrid has full exact exchange.  The first
term on the right of Eq.~(\ref{eq:eclh}) can be interpreted as
the G\"{o}rling--Levy limit of the static correlation term in
our local hybrid correlation of Eq.~(\ref{eq:lh}). (Only when
the static correlation energy arises from an \textit{exact}
degeneracy of the Kohn--Sham noninteracting system can the
corresponding  limit of the exact static correlation energy not
be finite~\cite{Perdew:2007/PRA/040501}.)

There are two regions of an atom or molecule in which $a_1\to 1$
is expected (and found by our local hybrid): the rapidly-varying
exponential density tail and the high-density region of the deep
core.  For the rapidly-varying density of a quantum well whose
thickness shrinks down to zero (the true two-dimensional limit),
$a_1\to 1$ is also expected (and found); our local hybrid improves
greatly but imperfectly~\cite{Constantin:2008/PRL/016406} upon TPSS
and other semilocal functionals in this limit.

In the slowly-varying limit, in which $s$ and
other dimensionless density derivatives tend to
zero, we have $\epsilon_\mathrm{c}^\mathrm{TPSS}\to
\epsilon_\mathrm{c}^\mathrm{LSD}$~\cite{Perdew:1996/PRL/3865,%
Tao:2003/PRL/146401,Perdew:1999/PRL/2544},
$\epsilon_\mathrm{c}^\mathrm{GL2TPSS}\to -\infty$
like $\ln s$ and $u\to\infty$, so $a_1\to 0$ like
$1/\ln(-\ln s)$ and $\epsilon_\mathrm{c}^\mathrm{lh}\to
\epsilon_\mathrm{c}^\mathrm{TPSS}$. Since the TPSS correlation
energy recovers~\cite{Perdew:1996/PRL/3865,Tao:2003/PRL/146401,%
Perdew:1999/PRL/2544} the second-order gradient
expansion~\cite{Ma:1968/PR/18} in this limit, so does our local
hybrid functional.

Finally, we note that in the many-electron low-density limit
[$\lambda\to 0$ under the uniform scaling of Eq.~(\ref{eq:uds})],
$1/\epsilon_\mathrm{x}^\mathrm{LSD}\to -\infty$, so $u\to\infty$,
$a_1\to 0$, and $\epsilon_\mathrm{xc}^\mathrm{lh}\to
\epsilon_\mathrm{xc}^\mathrm{TPSS}$. There are
reasons~\cite{Perdew:2004/JCP/6898} to believe that TPSS is accurate
(although not exact) for the exchange-correlation energy in this
limit: As correlation becomes stronger in comparison with exchange,
more error cancellation between TPSS exchange and TPSS correlation
is expected.

We have not been able to prove, for all possible densities,
that the local hybrid correlation energy is always negative or
that the Lieb--Oxford bound on its exchange-correlation energy is
always satisfied. These constraints are easier to guarantee with
semilocal functionals~\cite{Perdew:1996/PRL/3865,Tao:2003/PRL/146401,%
Perdew:1999/PRL/2544} than with local hybrids.

This $a_1$ makes our exchange-correlation functional exact in
the one-electron and high-density limits, and more correct in the
rapidly-varying limit. It should provide a $-1/r$ asymptote for
the effective exchange-correlation potential around an atom or
molecule in the limit $r\to\infty$, needed to bind small negative
ions~\cite{Perdew:1981/PRB/5048} and also needed in time-dependent
density functional theory~\cite{Book/Fiolhais:2003} when the electron
density is driven far from the nuclei. But it cannot deal with the
class of problems addressed in the next section.

\subsection{Abnormal regions where the hole density does not 
integrate to $-1$: Open subsystems of fluctuating electron number}
\label{subsec:4.2}

Semilocal approximations to the density functional for
the exchange-correlation energy assume that the exact
exchange-correlation hole density around an electron integrates
to $-1$ on the scale of the local Fermi wavelength, as it does
in an electron gas of slowly-varying density. Regions where
this is not true, e.g., open subsystems of fluctuating electron
number where $\Delta_\mathrm{xc}$ of Eq.~(\ref{eq:xc-hole-sum})
is nonzero~\cite{Perdew:1985/Dreizler/265,Perdew:2007/PRA/040501},
are therefore abnormal.

Consider the second interesting density- and position-dependent variable
\begin{equation}
 v = \frac{\epsilon_\mathrm{x}^\mathrm{ex}}
 {\epsilon_\mathrm{x}^\mathrm{TPSS}}.
\end{equation}
Where $v$ is sufficiently less than 1, we can
conclude [by the argument around Eqs.~(\ref{eq:nex})
and~(\ref{eq:xc-hole-sum})] that $\Delta_\mathrm{x}$ of
Eq.~(\ref{eq:sum-rule}) is positive.  Because correlation
suppresses density fluctuation~\cite{Ziesche:2000/IJQC/819},
we can expect $\Delta_\mathrm{xc} \leq \Delta_\mathrm{x}$.
We seek a mixing fraction $a_2$ such that $a_2\approx 1$ when
$\Delta_\mathrm{xc}\approx \Delta_\mathrm{x}\gg 0$, but $a_2\approx 0$
when $\Delta_\mathrm{xc}\approx 0$.

\begin{figure}
\includegraphics[width=\columnwidth]{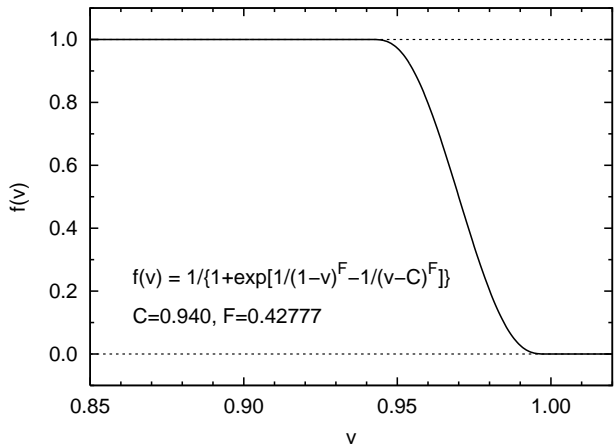}
\caption{\label{fig:fv}
Function $f(v)$ of Eq.~(\ref{eq:fv}).
The value of $C$ was found as explained in
Section~\ref{sec:fitting}. $F$ is related to $C$
by Eq.~(\ref{eq:F}).
}
\end{figure}

A first candidate for $a_2$ is
\begin{equation}
 f(v) = \left\{ \begin{array}{ll}
 1, & v \leq C \\
 \displaystyle{
 \frac{1}{1+\exp[1/(1-v)^{F} - 1/(v-C)^{F}]} }, & C< v< 1 \\
 0, & v \geq 1
 \end{array} \right.  \label{eq:fv}
\end{equation}
where $C\leq 1$ is an empirical parameter. We have found that we
can make $f(v)$ flat around $v=1$ or $C$, but not too steep at the midpoint
$v=(1+C)/2$, over a wide range of $C<1$, by choosing
\begin{equation}
 F = -\frac{3}{2\ln[(1-C)/2]} > 0.   \label{eq:F}
\end{equation}
The function $f(v)$ must be flat near $v=1$ since values of $v$
slightly less than 1 can arise even in normal regions.
It interpolates smoothly between 1 (at the highly abnormal $v<C$)
and 0 (at the normal $v\approx 1$), reaching $1/2$ at the midpoint
$v=(1+C)/2$, as illustrated in Fig.~\ref{fig:fv}.
If $C$ is close to 1, there is a sharp step in $f(v)$ which is
smoothed when $f(v)$ is multiplied by
$\epsilon_\mathrm{x}^\mathrm{TPSS}-\epsilon_\mathrm{x}^\mathrm{ex}
=(1-v)\epsilon_\mathrm{x}^\mathrm{TPSS}$.
In practical calculations, the function $f(v)$ is best evaluated
not by Eq.~(\ref{eq:fv}) but as described in Appendix~\ref{app:B}.

If $\Delta_\mathrm{xc}$ were always equal to $\Delta_\mathrm{x}$, we
could take $a_2=f(v)$. But the inequality $\Delta_\mathrm{xc}\leq
\Delta_\mathrm{x}$ leads us to consider simple examples in
which $v\ll 1$.  This situation arises in certain stretched-bond
diatomic molecules, with the bond length tending to infinity.

(a) In stretched H$_2^{+}$~\cite{Perdew:1998/Dobson/31}, a single
electron is shared equally between two well-separated protons as
H$^{+1/2}\cdots$H$^{+1/2}$. The electron fluctuates between the two
proton centers, making $\Delta_\mathrm{xc}=\Delta_\mathrm{x}>0$
on each center. The Hartree--Fock or exact-exchange-only energy
is correctly equal to the sum of the Hartree--Fock energies for
H and H$^{+}$, while the TPSS meta-GGA energy is 0.102 hartree
(64.0 kcal/mol) too low compared to the sum of the TPSS energies
for the integer-charge fragments. In this case, we want full exact
exchange or $a_2=1$.

(b) In stretched neutral
H$_2$~\cite{Perdew:2007/PRA/040501,Becke:2005/JCP/064101,%
Becke:2007/JCP/124108,Book/Levine:1991}, two electrons are
shared equally between two well-separated protons. In the
exact singlet ground-state wave function, there is always
exactly one electron on each proton center, because of static
correlation arising from a Kohn--Sham degeneracy which becomes
exact in the limit of infinite bond length, and each H ``atom"
is spin-unpolarized as a result of fluctuation between
``up" and ``down" spin, making $\Delta_\mathrm{xc}=0$ but
$\Delta_\mathrm{x}>0$.  The Hartree--Fock or exact-exchange-only
energy is incorrectly 0.285 hartree (179 kcal/mol) above the
Hartree-Fock energy of two atoms.  The corresponding TPSS error
is reduced to 0.083 hartree (51.8 kcal/mol).  In this case,
we want no exact exchange or $a_2=0$. We can define an average
$\bar{v}=E_\mathrm{x}^\mathrm{ex}/E_\mathrm{x}^\mathrm{TPSS}$, where
both $E_\mathrm{x}^\mathrm{ex}$ and $E_\mathrm{x}^\mathrm{TPSS}$
are evaluated with the converged TPSS orbitals. For a predominantly
normal system, $\bar{v}$ is close to 1 (e.g., 0.97 in H$_2^{+}$ and
0.99 in singlet H$_2$ at the respective equilibrium bond lengths),
but $\bar{v}$ is significantly less than 1 in abnormal systems (0.62
in infinitely-stretched H$_2^{+}$ and 0.61 in infinitely-stretched
singlet H$_2$).

\begin{figure}
\includegraphics[width=\columnwidth]{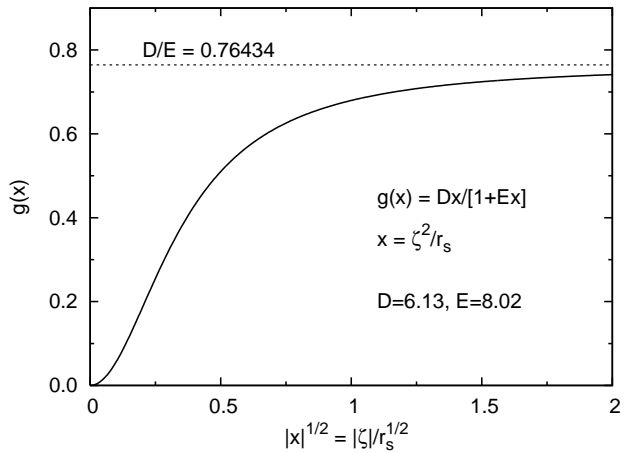}
\caption{\label{fig:gx}
Function $g(x)$ defined by Eq.~(\ref{eq:gx}).
The values of parameters $D$ and $E$ were fitted as explained in
Section~\ref{sec:fitting}.
}
\end{figure}

The hyper-GGA ingredient that seems to distinguish between these
two cases is the relative spin polarization
\begin{equation}
 \zeta = \frac{n_\uparrow-n_\downarrow}{n_\uparrow+n_\downarrow},
  \label{eq:zeta}
\end{equation}
which equals $\pm 1$ for stretched H$_2^{+}$ and 0 for stretched H$_2$.
Thus we try
\begin{equation}
 a_2 = g\left(\frac{\zeta^2}{r_s}\right) f(v),  \label{eq:a2}
\end{equation}
where $n=3/4\pi r_s^3$ and, setting $x=\zeta^2/r_s$,
\begin{equation}
 g(x) = \frac{Dx}{1+Ex},
  \label{eq:gx}
\end{equation}
in which $D$ and $E$ are positive empirical parameters, with $D\leq E$.
Eq.~(\ref{eq:gx}) is, like the exchange-correlation energy itself,
invariant under $\zeta\to -\zeta$.  It interpolates between 0
(at $x=0$) and $D/E$ (as $x\to\infty$), as shown in Fig.~\ref{fig:gx}
Our $a_2$ vanishes when
$\zeta\to 0$ (as in singlet H$_2$), but also when $r_s\to\infty$
(i.e., in the low-density limit where correlation can become
as strong as exchange, increasing the possibility of error
cancellation).  With this choice, it may be possible to get an
improvement in much of time-independent chemistry and condensed
matter physics from $a_2$ alone, although $a_1$ is still needed
for the situations described at the end of section~\ref{subsec:4.1}.
The selection of $\zeta$ as an ingredient
of $a_2$ is not altogether satisfactory, since it is motivated by
simple examples and not by general physical principles.

Our simple examples will now be discussed further: (i) In stretched
H$_2^{+}$, we have a one-electron density which is already properly
described by $a_1=1$ from section~\ref{subsec:4.1}. (ii) In stretched
H$_2$, the standard spin-symmetry breaking of semilocal functionals,
which has a good physical basis~\cite{Perdew:1995/PRA/4531}, will
localize a spin-up electron on one proton center, and a spin-down
electron on the other, making $|\zeta|=1$ almost everywhere and
eliminating fluctuations (making $\Delta_\mathrm{x}=0$).

Under the uniform density scaling of Eq.~(\ref{eq:uds}),
$v(\bfr)\to v(\lambda\bfr)$ and $\zeta(\bfr)\to\zeta(\lambda\bfr)$
and $r_s(\lambda\bfr)\to\lambda^{-1}r_s(\bfr)$.  [Similarly, for $s$
of Eq.~(\ref{eq:s}), $s(\bfr)\to s(\lambda\bfr)$].
In the high-density limit, our $a_2$ (like our $a_1$)
properly has an expansion in powers of $r_s$ (or $\lambda^{-1}$).

We have invoked five empirical parameters: two ($A$ and $B$) in $a_1$
and three ($C$, $D$, and $E$) in $a_2$. Next we will combine our
two mixing coefficients into one.

\subsection{Combination rule for the mixing fractions $a_1$ and $a_2$}
\label{subsec:4.3}

We must combine the mixing fractions $a_1$ and $a_2$ from
sections~\ref{subsec:4.1} and~\ref{subsec:4.2} to get a single
mixing fraction $a$ for use in Eq.~(\ref{eq:lh}), with the following
features: (i) $0\leq a \leq 1$, as expected for any local or global
hybrid, (ii) $a=1$ when \textit{either} $a_1=1$ or $a_2=1$, because
either condition indicates a strongly abnormal region in which full
exact exchange can be used. (iii) $a=a_2$ when $a_1=0$ and $a=a_1$
when $a_2=0$, since in either case all the abnormality of a region
is of either one type or the other. Condition (iii) also ensures
that $a=0$ when $a_1=a_2=0$, indicating a strongly normal region
in which semilocal approximations suffice.  More generally, $a >
\max(a_1,a_2)$, because $a$ should increase with abnormality.

A combination rule which satisfies all these expectations is
\begin{equation}
 a = 1 - (1-a_1)(1-a_2) = a_1 + a_2 - a_1 a_2.  \label{eq:a}
\end{equation}
Note that, in a nearly-normal region where $a_1$ and $a_2$ are small,
we find $a\approx a_1+a_2$. Combining $a_1$ with $a_2$ does not lose
any of the exact constraints satisfied by $a_1$ alone.  The flatness
of $f(v)$ at $v=1$ preserves the correct gradient expansion for
correlation in the slowly-varying limit. The $AB$ term of the
high-density ($\lambda\to\infty$) limit of Eq.~(\ref{eq:eclh}) is
multiplied by the finite $1-a_2(\lambda\bfr)$, and so remains finite.

\begin{figure}
\includegraphics[width=\columnwidth]{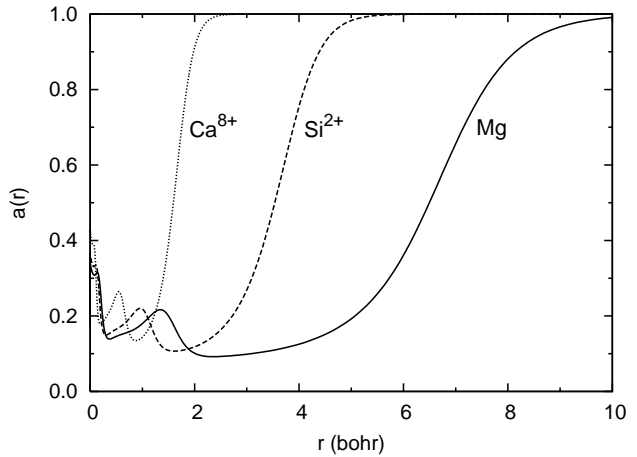}
\caption{\label{fig:mg}
PSTS hyper-GGA mixing fractions $a$ constructed for the Mg
atom and isoelectronic Si$^{2+}$ and Ca$^{8+}$ ions using
self-consistent TPSS orbitals obtained in Partridge's uncontracted
basis sets~\cite{Partridge:1987/JCP/6643,Partridge:1989/JCP/1043,%
Schuchardt:2007/JCIM/1045,Feller:1996/JCC/1571}:
(20s,12p) for Mg; (20s,15p) for Si; (23s,16p) for Ca.
The parameters of $a$ are given in Section~\ref{sec:fitting}.
Here $\zeta=0$, so $a_2=0$ and $a=a_1$. 
The figure illustrates that our mixing fraction $a$ tends smoothly
to 1 in the low-density tail, and slowly to 1 in the core in
the high-density ($Z\to\infty$) limit.
}
\end{figure}

\begin{figure}
\includegraphics[width=\columnwidth]{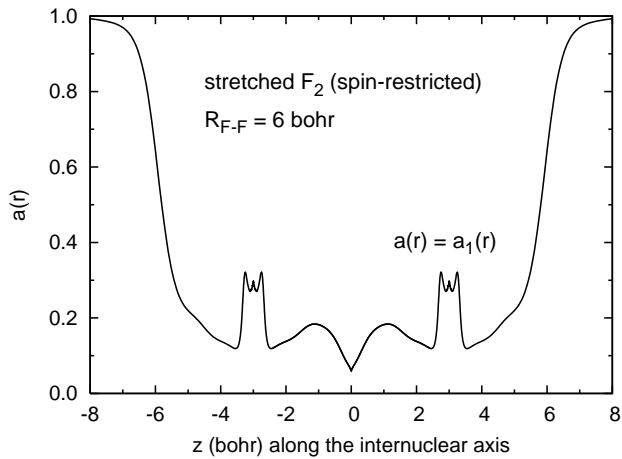}
\caption{\label{fig:f2}
PSTS hyper-GGA mixing fraction $a$ constructed with self-consistent
TPSS/u-6-311++G(3df,3pd) orbitals along the internuclear axis
of a stretched F$_2$ molecule ($z=0$ is the center of the bond).
In all regions shown, $a_2=0$ and $a=a_1$.
The parameters of $a$ are given in Section~\ref{sec:fitting}.
Note that $a$ has a cusp at $z=0$ which seems unphysical
but is limited to a very small region of three-dimensional space.
}
\end{figure}

\begin{figure}
\includegraphics[width=\columnwidth]{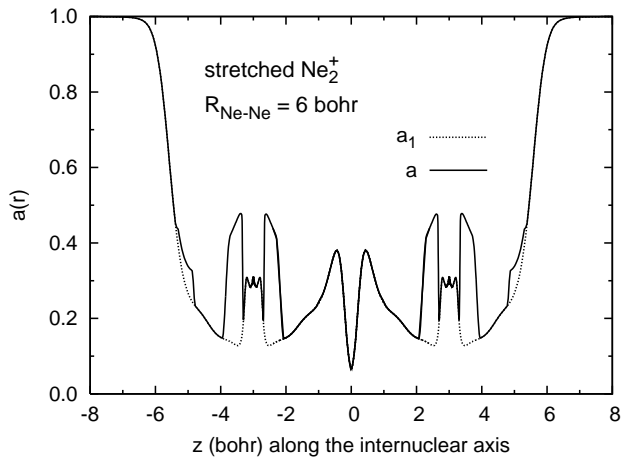}
\caption{\label{fig:ne2+}
PSTS hyper-GGA mixing fractions $a_1$ and $a$ 
constructed with self-consistent TPSS/u-6-311++G(3df,3pd) orbitals
along the internuclear axis of a stretched Ne$_2^{+}$ molecular ion
($z=0$ is the center of the bond).
The parameters of $a$ are given in Section~\ref{sec:fitting}.
In some regions, $a_2>0$ and $a>a_1$.
}
\end{figure}

For the convenience of having a name, we will call the constructed
local hybrid functional the PSTS hyper-GGA. The behavior of the
PSTS hyper-GGA mixing fraction $a$ and its components (with
parameters fitted as described in Section~\ref{sec:fitting})
is illustrated by Figs.~\ref{fig:mg}--\ref{fig:ne2+}. These
figures are qualitatively reasonable, as discussed near the end
of section~\ref{sec:conclusions}.

\section{Fitting the parameters to standard enthalpies of formation
and barrier heights}
\label{sec:fitting}

We have determined the five empirical parameters $(A,B,D,C,E)$
of the PSTS hyper-GGA mixing fraction $a(\bfr)$ by minimizing the quantity
\begin{equation}
 S_W = \frac{1}{2} \left[ \mbox{MAE(G2/97)}
 + \mbox{MAE(BH42/03)} \right],   \label{eq:SW}
\end{equation}
where MAE(G2/97) is the calculated mean absolute error of 148
standard enthalpies of formation of the molecules of the G2/97
test set~\cite{Curtiss:1997/JCP/1063}, and MAE(BH42/03) is the
mean absolute error of 42 forward and reverse barrier heights of
the 21 gas-phase hydrogen-transfer reactions of the BH42/03 test
set~\cite{Zhao:2004/JPCA/2715}. The molecules of the G2/97 set are
built up from hydrogen and/or first- and second-row atoms, and all
but one of the reactions of the BH42/03 test set involve atomic
and molecular radicals (i.e., spin-unpaired systems) as reactants.
The reference values of the standard enthalpies of formation are
all experimental, while the 42 barriers are best estimates based
on a combination of experimental reaction rates and high-level
electronic structure calculations~\cite{Zhao:2004/JPCA/2715}.
The mean signed error of standard enthalpies of formation
is roughly equal and opposite to the mean signed error of
the corresponding atomization energies, whereas the mean
absolute errors are about the same. The rather complicated
methodology of evaluating the standard enthalpies of formation is
documented in Ref.~\onlinecite{Staroverov:2003/JCP/12129}. The
only difference between the present procedure and the one of
Ref.~\onlinecite{Staroverov:2003/JCP/12129} is that here we employed
the fully uncontracted 6-311++G(3df,3pd) basis set for calculating
the electronic energies rather than the standard (contracted)
6-311++G(3df,3pd) basis~\cite{Schuchardt:2007/JCIM/1045}, because
the latter is not well-suited for the resolution of the identity
used for evaluating $\epsilon_\mathrm{x}^\mathrm{ex(conv)}$ and
$\epsilon_\mathrm{x}^\mathrm{ex}$~\cite{Tao:2008/PRA/012509}.

\begin{table*}
\caption{\label{table:1}
Relative performance of the PSTS and two simplified hyper-GGAs
for the G2/97 and G3/99 test sets of standard enthalpies
of formation (148 and 223 molecules, respectively) and the BH42/03
set of 42 reaction barriers. The test sets are described in
Section~\ref{sec:fitting}. G2/97 is a subset of G3/99.
ME is the mean error (signed) relative to reference values,
MAE is the mean absolute error. $S_W$ is defined by Eq.~(\ref{eq:SW}).
All calculations were carried out with a development version
of the \textsc{gaussian} program~\cite{GDV-F02} using the fully
uncontracted 6-311++G(3df,3pd) basis set. All values are in kcal/mol.
For comparison, the mean atomization energies are 478.5 kcal/mol
for G2/97 and 1180 kcal/mol for G3/99, while the mean reaction barrier
height is 14.0 kcal/mol (BH42/03).
$\mbox{1 kcal/mol} \approx \mbox{0.0434 eV} \approx \mbox{0.00159 hartree}$.
}
\begin{ruledtabular}
\begin{tabular}{lrrrrrrrrr}
 & \multicolumn{2}{c}{G2/97} & \multicolumn{2}{c}{BH42/03}
 &  & \multicolumn{4}{c}{G3/99} \\
  \cline{2-3} \cline{4-5} \cline{7-10}
 Method & ME & MAE & ME & MAE & $S_W$
  & ME & MAE & Max. ($+$) & Min. ($-$) \\ \hline
Hartree--Fock & 148.0 & 148.0 & 12.5 & 12.8 & 80.4 &
 211.0 & 211.0 & 581.0 (C$_8$H$_{18}$) & $-$0.6 (BeH) \\
HFx/TPSSc & 25.2 & 27.7 & 4.2 & 5.8 & 16.7 &
 27.7 & 30.3 & 130.0 (O$_3$) & $-$23.1 (Si$_2$H$_6$) \\
LSDA\footnotemark[1] & $-$83.4 & 83.4 & $-$18.1 & 18.1 & 50.7 &
 $-$121.2 & 121.2 & 0.6 (Li$_2$) & $-$344.3 (C$_{10}$H$_8$) \\
PBE GGA & $-$16.1 & 16.9 & $-$9.7 & 9.7 & 13.3 &
 $-$21.6 & 22.1 & 10.8 (Si$_2$H$_6$) & $-$78.9 (C$_{10}$H$_8$) \\
TPSS meta-GGA & $-$5.7 & 6.4 & $-$8.4 & 8.4 & 7.4 &
  $-$6.0 & 6.5 & 15.0 (SiF$_4$) & $-$23.9 (ClF$_3$) \\
PBEh & $-$2.6 & 5.0 & $-$4.7 & 4.7 & 4.8 &
 $-$5.0 & 6.8 & 20.5 (SiF$_4$) & $-$35.8 (C$_{10}$H$_8$) \\
TPSSh & $-$1.9 & 4.4 & $-$6.6 & 6.6 & 5.5 &
  $-$1.6 & 4.1 & 20.8 (SiF$_4$) & $-$18.0 (Si$_2$H$_6$) \\
%B3LYP & 0.9 & 3.1 & $-$4.7 & 4.7 & 3.9 &
% 3.3 & 4.8 & 19.3 (SiF$_4$) & $-$8.0 (BeH) \\
PSTS hyper-GGA\footnotemark[2] & $-$0.6 & 4.7 & $-$1.2 & 2.1 & 3.4 &
 $-$0.2 & 4.5 & 23.2 (SiF$_4$) & $-$19.1 (Si$_2$H$_6$) \\
Hyper-GGA (conv.)\footnotemark[2]
 & $-$0.8 & 5.6 & $-$1.2 & 2.3 & 3.9 &
 $-$0.2 & 5.5 & 28.7 (SiF$_4$) & $-$21.5 (Si$_2$H$_6$) \\
Hyper-GGA ($a=a_1$)\footnotemark[2]
 & 0.2 & 4.5 & $-$6.6 & 6.6 & 5.5 &
 1.5 & 4.7 & 26.5 (SiF$_4$) & $-$19.0 (Si$_2$H$_6$) \\
\end{tabular}
\end{ruledtabular}
\footnotetext[1]{Using the Perdew--Wang representation~\cite{Perdew:1992/PRB/13244}
of $\epsilon_\mathrm{c}^\mathrm{LSDA}(r_s,\zeta)$.}
\footnotetext[2]{All hyper-GGA energies were evaluated using self-consistent
TPSS orbitals.}
\end{table*}

We do not have yet a fully self-consistent implementation of
our local hybrid functional and so evaluate all hyper-GGA energies
using converged TPSS orbitals. We speculate, based on our studies
of a previous version of the hyper-GGA, that the choice of orbitals
would have a small effect on the results for enthalpies of formation
and barrier heights, provided that the parameters of the functional
are optimized for each choice.

The optimized values of the hyper-GGA parameters obtained in this
manner are as follows: $A=2.74$, $B=132$, $C=0.940$, $D=6.13$,
and $E=8.02$. Although the G2/97 and BH42/03 training sets differ
markedly in size, their equal weighting in Eq.~(\ref{eq:SW})
does not cause a significant bias in favor of the data included
in either test set because both sets are sufficiently representative
of the types of molecules/reactions they contain.

In Table~\ref{table:1}, we compare the performance of our PSTS
hyper-GGA with that of simpler functionals constructed by the method
of constraint satisfaction with no or minimal empiricism. The
functionals not already introduced in Section~\ref{sec:intro}
are: HFx/TPSSc is the Hartree--Fock exchange with TPSS
correlation, PBEh is the global hybrid PBE functional with
$a=0.25$~\cite{Perdew:1996/JCP/9982,Ernzerhof:1999/JCP/5029,%
Adamo:1999/JCP/6158}, and TPSSh is the global TPSS hybrid with
$a=0.10$~\cite{Staroverov:2003/JCP/12129}. Among the approximations
included in Table~\ref{table:1}, the PSTS hyper-GGA is overall the
most accurate (has the smallest $S_W$). To put these results into
perspective, we point out that the mean experimental atomization
energy for the G2/97 set of molecules is 478.5 kcal/mol, while
the mean experimental barrier height for the BH42 set of reaction
barriers is 14.0 kcal/mol. Note that although the PSTS hyper-GGA
has been trained on G2/97, it has an even better performance for the
larger G3/99 test set~\cite{Curtiss:2000/JCP/7374} of 223 molecules
(including COF$_2$), for which the mean experimental atomization
energy is $\sim$1180 kcal/mol.

Table~\ref{table:1} also shows results for two simplified
hyper-GGAs labeled ``hyper-GGA (conv.)" and ``hyper-GGA
($a=a_1$)". The former differs from the PSTS hyper-GGA in
that it uses the conventional exact-exchange energy per
electron $\epsilon_\mathrm{x}^\mathrm{ex(conv)}$ in place of
the gauge-transformed $\epsilon_\mathrm{x}^\mathrm{ex}$ of
Eq.~(\ref{eq:exg}). The five empirical parameters of this form,
determined by minimizing $S_W$, are as follows: $A=3.14$, $B=146$,
$C=0.930$, $D=5.17$, and $E=9.49$. This form performs slightly
worse than the PSTS hyper-GGA, which is consistent with our
argument that local hybrids should combine exact and semi-local
exchange in the same gauge. The second simplified hyper-GGA
uses $\epsilon_\mathrm{x}^\mathrm{ex}$ in the TPSS gauge but has
$a_2$ set to zero, so its mixing fraction $a=a_1$ depends only
on parameters $A$ and $B$ (whose optimized values are $A=2.77$
and $B=545$). Because this $a$ is close to 0.1 for nearly all
relevant values of $u(\bfr)$, the performance of the $a_1$-only
hyper-GGA is very similar to that of TPSSh (a global hybrid with
$a=\mbox{const}=0.1$): atomization energies are more accurate
than from the TPSS meta-GGA, but there is little improvement for
reaction barriers.

\section{Conclusions}
\label{sec:conclusions}

We have argued that semilocal density functionals work in many
cases because of a justified cancellation of errors between
exchange and correlation that occurs in identifiable ``normal"
regions. This enables one to construct nonempirical semilocal
functionals (such as LSD, PBE GGA, and TPSS meta-GGA) on the
first three rungs of a ladder of increasingly sophisticated
approximations. The fourth or hyper-GGA rung requires empirical
parameters to balance the full nonlocality of correlation against
that of exact exchange, since known exact constraints say nothing
about this balance. (At least this is so when we consider only
exact constraints on the integrated exchange-correlation energy;
a correlation factor~\cite{Gori-Giorgi:2002/PRB/165118}, applied
to a nearly-exact exchange hole density of an inhomogeneous
system~\cite{Ernzerhof:2008/talk}, might work or might not
without empiricism). Our PSTS hyper-GGA is a local hybrid,
mixing in a fraction $a$ of exact exchange locally, according to
Eqs.~(\ref{eq:lh}), (\ref{eq:a1}), (\ref{eq:a2}), and (\ref{eq:a}).
Because our $a$ tends to one in the high-density limit, our
hyper-GGA has full exact exchange. It is also one-electron
self-interaction-free and size-consistent.

The small relative error (typically of order 1\% or less) of the TPSS
meta-GGA for the atomization energies of molecules suggests that
these quantities are dominated by contributions from normal regions.
The much larger relative error (typically of order 60\%) of the TPSS
meta-GGA for the barrier heights to chemical reactions suggests that
those quantities contain substantial contributions from abnormal
regions. Thus we fit our hyper-GGA parameters simultaneously to
atomization energies (preserving and slightly improving them,
mainly via our local mixing fraction $a_1$) and  to barrier heights
(substantially improving them, mainly through our local mixing
fraction $a_2$).

We have invoked five empirical parameters $(A,B,C,D,E)$, probably
close to the minimum possible number since there are four kinds of
abnormal regions.  But, as more empirical parameters are introduced,
there is a graver danger of ``overfitting" any given limited data
set. A fit that is unjustifiably good can worsen results for systems
and properties very different from those that have been fitted. We
have tried to minimize this danger by fitting to forms which
take into account known exact constraints, physical insights, and
paradigm examples.

To construct our PSTS hyper-GGA, we satisfy exact constraints
on the density functional for the exchange-correlation energy.
Of all the constraints possible for a hyper-GGA, the only one we have not
tried to satisfy is the non-zero limit~\cite{Perdew:1999/PRL/2544}
for the correlation energy under one-dimensional density scaling to
the true two-dimensional limit.  In this limit, our PSTS hyper-GGA
correlation energy tends to zero (as in the TPSS meta-GGA); the
correct non-zero limit seems too complicated to incorporate in any
simple way, and not very relevant to most physical systems. Moreover,
two exact constraints that are guaranteed for all possible densities
at the TPSS meta-GGA level are no longer so guaranteed in local
hybrids like PSTS: the non-positivity of the correlation energy and
the Lieb--Oxford lower bound on the exchange-correlation energy. Of
course, the PSTS exchange-correlation energy is negative for all
possible densities.

Apart from the need for empirical parameters, constraint
satisfaction is the same method
used earlier to construct the PBE GGA~\cite{Perdew:1996/PRL/3865}
and the TPSS meta-GGA~\cite{Tao:2003/PRL/146401}. In the
hyper-GGA case, we satisfy additional exact constraints via
a careful interpolation between a semilocal exchange energy
density in a normal region and the exact exchange energy density
in an abnormal region.  Our method combines (and goes beyond)
some of the best formal features of other methods, including the
size-consistency of methods that model the exchange-correlation
hole density~\cite{Becke:2005/JCP/064101,Becke:2007/JCP/124108,%
Becke:1989/PRA/3761,Perdew:1996/PRB/16533} and the proper
concern for the high-density (or weakly-interacting)
and low-density (or strongly-interacting) limits of
methods that model the integrand $W_\alpha[n]$ of the
coupling-constant integral~\cite{Mori-Sanchez:2006/JCP/091102,%
Cohen:2007/JCP/034101,Becke:1993/JCP/5648,Perdew:1996/JCP/9982,%
Seidl:2000/PRL/5070}
\begin{equation}
  E_\mathrm{xc}[n]=\int_0^1 d\alpha\, W_\alpha[n].
\end{equation}
Of course, our PSTS hyper-GGA has its own hole model
[namely, Eq.~(\ref{eq:nxc})] and its own coupling-constant
integrand~\cite{Gorling:1993/PRB/13105,Levy:1985/PRA/2010}
defined by
\begin{equation}
 W_\alpha[n] = \frac{d}{d\alpha} (\alpha^2 E_\mathrm{xc}[n_{1/\alpha}]).
  \label{eq:W-alpha}
\end{equation}

The behavior of our abnormality index or local
exact-exchange mixing fraction $a(\bfr)$, as illustrated in
Figs.~\ref{fig:mg}--\ref{fig:ne2+}, is qualitatively reasonable:
For an isolated many-electron atom of fixed electron number
(Fig.~\ref{fig:mg}) it is small ($\approx 0.1$) in the
valence region but larger ($\approx 0.3$) in the higher-density core,
and it gradually approaches 1 in the rapidly-varying exponential
tail. For a stretched molecule, it is atomic-like around an atom
of weakly-fluctuating electron number (Fig.~\ref{fig:f2} for
spin-restricted stretched F$_2$), but much larger ($\approx 0.4$)
in the relevant valence region of an atom of strongly-fluctuating
electron number (Fig.~\ref{fig:ne2+} for stretched Ne$_2^{+}$).
A version of the last effect is what raises and improves
the barrier heights in Table~\ref{table:1}.

Figure~\ref{fig:mg} shows a slow increase of $a$ in the core as the
atomic number $Z$ increases from 12 (Mg) to 14 (Si$^{2+}$) to 20
(Ca$^{8+}$). At large $Z$, there is a uniform density scaling to the
high-density limit in the core, so that $a\to 1$ as $Z\to\infty$. The
slowness of the approach to 1 is not a problem, since TPSS meta-GGA
(unlike LSDA) exchange is accurate for localized core-electron
densities.

Figure~\ref{fig:f2} shows a seemingly unphysical cusp in $a$ at the
bond center, where for the symmetric stretched molecule F$_2$ the
reduced density gradient $s\to 0$ and thus our $a\to 0$ (although the
approach to 0 cannot be plotted on the scale of this figure). This
cusp is a consequence of our Eqs.~(\ref{eq:u}) and (\ref{eq:a1}),
since $u\to\infty$ as $s\to 0$. By recovering semilocality in the
limit of slowly-varying density, we have introduced an artifact
in a small volume around the bond center. Stretched Ne$_2^{+}$
in Fig.~\ref{fig:ne2+} also displays a bond-center cusp.
Even the TPSS meta-GGA exchange by itself can have a bond-center
artifact~\cite{Perdew:2004/JCP/6898}. So far as we know, these
artifacts are harmless.

The ideas in this paper are general ones that may be generally
useful. On the other hand, our way of implementing them is far
from unique. In future work, we hope to continue to refine and
test this approach to the hyper-GGA including its self-consistent
implementation.

\appendix
\section{Expression for $\epsilon_\mathrm{c}^\mathrm{GL2TPSS}$}
\label{app:A}

The G\"{o}rling--Levy second-order limit of the TPSS correlation energy
per electron $\epsilon_\mathrm{c}^\text{GL2TPSS}$ is given by
\begin{equation}
 \epsilon_\mathrm{c}^\text{GL2TPSS}
 = \epsilon_\mathrm{c}^\text{GL2revPKZB} \left[
 1 + d \epsilon_\mathrm{c}^\text{GL2revPKZB}
 \left( \frac{\tau_W}{\tau} \right)^3 \right],
\end{equation}
where $d=2.8$ hartree$^{-1}$ is a constant and
\begin{eqnarray}
 & & \hspace*{-0.5cm}
 \epsilon_\mathrm{c}^\text{GL2revPKZB} \nonumber \\
 & & = \epsilon_\mathrm{c}^\mathrm{GL2PBE}(n_\uparrow,n_\downarrow,
 \nabla n_\uparrow,\nabla n_\downarrow) \left[
 1 + C(\zeta,\xi) \left( \frac{\tau_W}{\tau} \right)^2 \right]
 \nonumber \\
 & & - \left[ 1 + C(\zeta,\xi) \right]
 \left( \frac{\tau_W}{\tau} \right)^2
 \sum_\sigma \frac{n_\sigma}{n}
 \tilde{\epsilon}_\mathrm{c,\sigma}^\mathrm{GL2PBE}.  \label{eq:GL2revPKZB}
\end{eqnarray}
In Eq.~(\ref{eq:GL2revPKZB}), $\epsilon_\mathrm{c}^\text{GL2PBE}$
is the G\"{o}rling--Levy limit of the PBE correlation
energy per electron. It is obtained by replacing $\lambda\bfr$ with
$\bfr$ in the $\lambda\to\infty$ uniform density scaling limit
of the PBE correlation energy per electron~\cite{Perdew:1996/PRL/3865}:
\begin{eqnarray}
 & & \hspace*{-0.5cm}
 \epsilon_\mathrm{c}^\text{GL2PBE}
 (n_\uparrow,n_\downarrow,\nabla n_\uparrow,\nabla n_\downarrow)
 \nonumber \\
 & & = - \gamma\phi^3 \ln \left[
 1 + \frac{1}{\chi s^2/\phi^2 + (\chi s^2/\phi^2)^2 } \right],
\end{eqnarray}
in which $\gamma=(1-\ln 2)/\pi^2$,
\begin{equation}
 \phi(\zeta) = \frac{1}{2} \left[
 (1+\zeta)^{2/3} + (1-\zeta)^{2/3} \right],
\end{equation}
$s=|\nabla n|/2nk_F$ is the reduced density gradient of
Eq.~(\ref{eq:s}), $k_F=(3\pi^2n)^{1/3}$,
and $\chi=(\beta/\gamma) c^2 e^{-\omega/\gamma} \approx 0.72161$,
where $c = (3\pi^2/16)^{1/3}$, $\beta=0.066725$, and
$\omega=0.046644$~\cite{Perdew:1996/PRL/3865}.
The spin-dependent function
$\tilde{\epsilon}_\mathrm{c,\sigma}^\mathrm{GL2PBE}$ is
defined~\cite{Tao:2003/PRL/146401,Perdew:2004/JCP/6898} as
\begin{eqnarray}
 \tilde{\epsilon}_\mathrm{c,\sigma}^\mathrm{GL2PBE}
 & = & \max \big[ \epsilon_\mathrm{c}^\mathrm{GL2PBE}
 (n_\sigma,0,\nabla n_\sigma,0), \nonumber \\
 & & \epsilon_\mathrm{c}^\mathrm{GL2PBE}(n_\uparrow,n_\downarrow,
 \nabla n_\uparrow,\nabla n_\downarrow) \big].
\end{eqnarray}
The function $C(\zeta,\xi)$, where $\zeta$
is the spin-polarization of Eq.~(\ref{eq:zeta}) and
$\xi=|\nabla\zeta|/2k_F$, is given by Eq.~(14) of
Ref.~\onlinecite{Tao:2003/PRL/146401}.

\section{Evaluation of $f(v)$}
\label{app:B}

The exponent $1/(1-v)^F-1/(v-C)^F$ appearing in Eq.~(\ref{eq:fv})
tends to $+\infty$ as $v$ approaches 1 from below. In order to avoid
numerical overflow problems with the exponential term of $f(v)$ in
the interval $C<v<1$, Eq.~(\ref{eq:fv}) may be rewritten as follows. Define
\begin{equation}
 p_1 = \frac{1}{(1-v)^F}, \quad
 p_C = \frac{1}{(v-C)^F}.
\end{equation}
Then, inside the interval $C<v<1$,
\begin{equation}
 f(v) = \left\{ \begin{array}{ll}
 \displaystyle{\frac{1}{1+e^{p_1-p_C}}}, & \mbox{if $p_1\leq p_C$} \\
 \displaystyle{\frac{e^{p_C-p_1}}{1+e^{p_C-p_1}}}, & \mbox{if $p_C\leq p_1$} \\
 \end{array} \right., \label{eq:fv1}
\end{equation}
which is robust because all exponents are non-positive.

\begin{acknowledgments}
This work was supported by the National Science Foundation (NSF)
under Grants DMR-0501588 (J.P.P.) and CHE-0807194 (G.E.S.), by the
Natural Sciences and Engineering Research Council of Canada (NSERC)
through the Discovery Grants Program (V.N.S.), and by the Department
of Energy under Grant No. LDRD-PRD X9KU at LANL (J.T.)
\end{acknowledgments}

\end{document}